Article

# Emergent exchange-driven giant magnetoelastic coupling in a correlated itinerant ferromagnet

Check for updates

Carolina A. Marques [1] ✉, Luke C. Rhodes [1], Weronika Osmolska [1], Harry Lane[1], Izidor Benedičič [1], Masahiro Naritsuka [1], Siri A. Berge [1], Rosalba Fittipaldi [2], Mariateresa Lettieri [2], Antonio Vecchione [2] & Peter Wahl [1,3] ✉

The interaction between the electronic and structural degrees of freedom is central to several intriguing phenomena observed in condensed-matter physics. In magnetic materials, magnetic interactions couple to lattice degrees of freedom, resulting in magnetoelastic coupling, which is typically small and only detectable in macroscopic samples. Here we demonstrate a giant magnetoelastic coupling in the correlated itinerant ferromagnet $Sr_4Ru_3O_{10}$. We establish an effective control of magnetism in the surface layer and utilize it to probe the impact of magnetism on its electronic and structural properties. By using scanning tunnelling microscopy, we reveal subtle changes in the electronic structure dependent on ferromagnetic or antiferromagnetic alignment between the surface and subsurface layers. We further determine the consequences of the exchange force on the relaxation of the surface layer, which exhibits giant magnetostriction. Our results provide a direct measurement of the impact of exchange interactions and correlations on structural details in a quantum material, revealing how electronic correlations result in a strong electron–lattice coupling.

The interplay between electronic, magnetic and lattice degrees of freedom leads to rich phase diagrams in strongly correlated electron materials. When an external parameter is used to modify one of these degrees of freedom, the others are inadvertently affected. Examples include the emergence of and relationship between charge density wave orders and superconductivity in cuprate high-temperature superconductors[1–6], colossal magnetoresistance in manganites[7] and ruthenates[8,9], and magnetostriction in itinerant metamagnetic materials.[10,11] Coulomb repulsion and the Pauli principle drive electron correlations. To understand the properties of these materials and enable designing them with specific ground states requires understanding the impact of correlations on the coupling between electronic, magnetic and lattice degrees of freedom and their impact on macroscopic properties. In materials with the additional strong spin–orbit coupling, the interplay between magnetic and electronic degrees of freedom can result in a field-tunable electronic structure, as seen in kagome materials[12] and $Sr_3Ru_2O_7$ (ref. [13]), and Weyl nodes that can be controlled through magnetism and structural distortions.[14] Here we establish the impact of exchange interactions on structural properties from atomic-scale imaging and spectroscopy.

If spin–orbit coupling can be neglected, the coupling between magnetic and lattice degrees of freedom is dominated by exchange interactions. This can be illustrated by considering two electrons centred at neighbouring sites: the Pauli principle requires a spatially antisymmetric wavefunction if the spins are parallel (Fig. 1a), whereas the wavefunction has to be symmetric for antiparallel alignment of the spins (Fig. 1b). The exchange interaction arises from the change in overlap of the wavefunctions between the two configurations, resulting in different amounts of Coulomb repulsion and hence, different ground-state energies. The exchange energy depends sensitively on the distance between the two atoms. For elemental solids, the distance dependence is described by the Bethe–Slater curve[15,16], with an antiferromagnetic ground state for small interatomic distances, and a ferromagnetic ground state at intermediate and larger distances at which the interaction decays (Fig. 1c). The Bethe–Slater curve explains the magnetic ground states of elemental metals based on their lattice constant and provides a simple description of the magnetostriction in materials in which the spin–orbit coupling is negligible, for example, in nickel[17]. Similar arguments as for simple metals based on the Bethe–Slater curve apply for interactions in more complex materials, such as magnetic order in monoatomic chains[18], exchange interaction in amorphous metallic alloys[19] and interlayer exchange interaction in two-dimensional (2D) van der Waals ferromagnets[20]. The interaction also results in an exchange force: in the regime of decaying interaction, an antiparallel alignment results in an outward force compared with the parallel alignment of magnetization, where the exchange interaction pulls the atoms closer to each other. Such exchange magnetostriction has recently been experimentally confirmed in mechanical resonators built using 2D van der Waals ferromagnets[21]. Furthermore, in bilayers of van der Waals materials, a subtle interplay between the structure and magnetic ground state has been seen[22]. In transition metal compounds, the magnetostriction typically remains well below 100 ppm, unless a strong spin–orbit coupling plays a role[23]; larger values can be considered as 'giant'. Here we demonstrate the direct detection of exchange magnetostriction for the strongly correlated itinerant ferromagnet $Sr_4Ru_3O_{10}$.

[1]SUPA, School of Physics and Astronomy, University of St Andrews, North Haugh, St Andrews, UK. [2]CNR-SPIN, c/o Università di Salerno, Fisciano (SA), Italy. [3]Physikalisches Institut, Universität Bonn, Bonn, Germany. ✉e-mail: cdamarques@outlook.com; wahl@st-andrews.ac.uk

$Sr_4Ru_3O_{10}$ is a metamagnetic material that undergoes a ferromagnetic transition at $T_C \approx 100$ K (refs. 24,25) and a further change in the ordered moment below $T^* \approx 50$ K (ref. 26). The material shows substantial magnetostructural coupling: the magnetic order is highly sensitive to pressure, which results in the canting of spins[27], and the bulk material exhibits sizeable magnetostriction[28,29]. $Sr_4Ru_3O_{10}$ has a strongly anisotropic, quasi-2D electronic structure with a negligible out-of-plane coupling[30], suggesting that the magnetism also behaves in a quasi-2D manner. Previous studies by low-temperature scanning tunnelling microscopy (STM) have provided detailed information about the low-energy electronic structure[29] and show evidence for multiple Van Hove singularities (VHSs) in the vicinity of the Fermi energy[30]. We use ultralow-temperature STM to investigate the interplay of magnetic, electronic and structural degrees of freedom at the surface of $Sr_4Ru_3O_{10}$. In spin-polarized STM, the intensity of the differential conductance can be used to probe the change in the magnetic ground state of an adatom as a function of its distance to a ferromagnetic monolayer[31]. Here, using non-spin-polarized STM, we show that the energy of the VHSs can be used as a probe for the relative orientation of the surface and bulk magnetizations, allowing us to extract the magnetostructural changes associated with different magnetic configurations with sub-picometre precision. Our measurements reveal a giant exchange magnetostriction in $Sr_4Ru_3O_{10}$ and the 2D nature of its magnetism.

## Results

### Detecting the interlayer coupling

Since the magnetic ground state of $Sr_4Ru_3O_{10}$ is ferromagnetic, the magnetizations of surface layer ($M_S$) and bulk ($M_B$) are expected to be aligned ferromagnetically (Fig. 2a, left) after zero-field cooling or after polarizing the sample magnetization in high magnetic fields and ramping back to 0 T (Supplementary Section 1). In this case, the tunnelling spectrum (Fig. 2b, black curve) shows two peaks just below $E_F$, namely, −2.6 mV ($P_I$) and −0.8 mV ($P_{II}$)[30].

For magnetic fields larger than the coercive field of the bulk ($B_B$) but applied in the opposite direction of the bulk magnetization $M_B$, $M_B$ will flip direction. If $M_S$ does not flip at the same field as the bulk, this results in a metastable antiferromagnetic alignment (Fig. 2a, right). As a consequence, we expect subtle structural changes, as predicted by density functional theory (DFT) calculations (Supplementary Section 2). We also expect small changes in the electronic structure due to a finite interlayer coupling, which can be detected in the tunnelling spectra (Supplementary Section 3). Indeed, at magnetic fields slightly above $B_B$ in the opposite direction as the field with which the sample was polarized, we find that the energy of peak $P_I$ increases from −2.6 mV to −1.85 mV, whereas $P_{II}$ remains at a constant energy of −0.8 mV (Fig. 2b, purple curve).

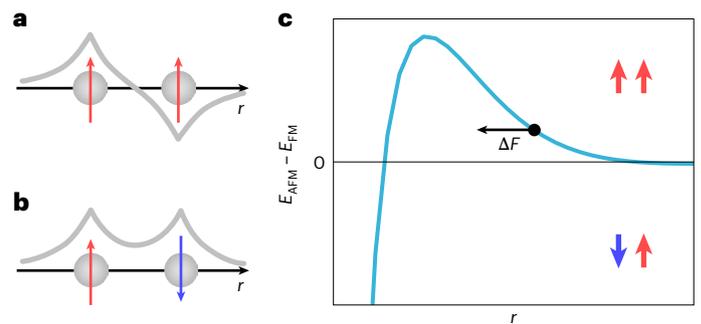

**Fig. 1 | Bethe–Slater-type exchange interaction. a,b**, Illustration of the symmetry of the wavefunction of two electrons at neighbouring sites, being spatially antisymmetric for parallel spins (**a**) and symmetric for antiparallel spins (**b**). **c**, Sketch of the Bethe–Slater curve showing the energy difference between the antiferromagnetic ($E_{AFM}$) and ferromagnetic ($E_{FM}$) configurations of a system of two electrons as a function of the distance ($r$) between them. The black arrow indicates the direction of the exchange force ($F$) for a system in the decaying part of the curve (black circle): the exchange force points towards lower $r$ if the spins are parallel, whereas it points towards larger $r$ if the spins are antiparallel.

Figure 2c shows the detailed field dependence of the spectra taken at 80 mK, which reveals two sudden changes: starting from a positive field with the surface and bulk polarized in a field of $B_z = +1.3$ T along the $c$ direction, the first change occurs between −0.13 T and −0.17 T and then the second one between −0.96 T and −1.06 T. At the two jumps, the energy of $P_I$ changes, moving up in energy by 750 μV before jumping down again. The bulk coercive field of $Sr_4Ru_3O_{10}$ is close to 0.2 T (Supplementary Section 4), suggesting that the jump between −0.13 T and −0.17 T coincides with the magnetization of the bulk reversing to align with the applied field. The detailed field dependence demonstrates that the VHSs, $P_I$ and $P_{II}$, can be used as indicators for the relative magnetizations of the surface and subsurface layers.

Despite the small fields of less than 1 T used here, the high energy resolution of our tunnelling spectra allows us to determine the spin character of the VHS $P_I$, providing independent confirmation of the magnetization of the surface layer. From measurements at high fields, we can determine that both $P_I$ and $P_{II}$ are spin polarized and have a spin-majority character, since both move towards lower energies with an increasing magnetic field (Supplementary Fig. 5). Figure 2d shows the peak position of $P_I$ as a function of the magnetic field for a full magnetic field loop starting from a loop field of $B_L = +1.3$ T, ramping to −1.06 T and back to +1.05 T. For both field directions, we detect a clear jump at $B_z \approx 0.15$ T, where the bulk switches magnetization as the surface magnetization remains aligned antiparallel with the field. The second jump at $B_z \approx \pm 1$ T occurs when the surface magnetization switches and aligns parallel with the applied field and bulk magnetization. The field-induced shift in $P_I$ between these two jumps reflects the magnetization direction of the surface layer: initially, although the magnetization of the surface is parallel to the field and for field $B_z$ smaller than the bulk coercive field $B_B$ ($B_z < B_B$), $P_I$ shifts to more negative energies with

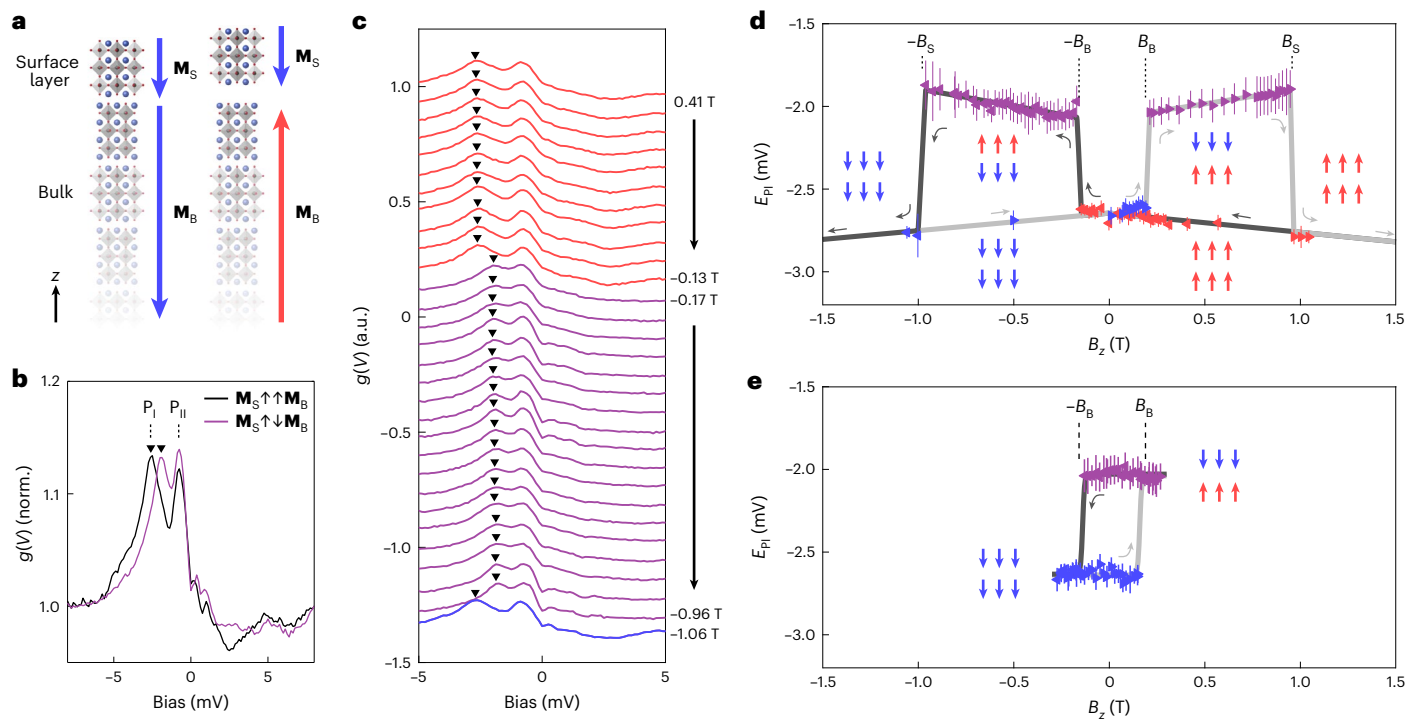

**Fig. 2 | Detecting surface magnetization from the tunnelling spectra. a**, Sketch of the crystal structure of the top layers and magnetization of the bulk ($M_B$) and surface layer ($M_S$). In the ferromagnetic ground state[26], both $M_S$ and $M_B$ point in the same direction parallel to the $c$ axis (left). When the direction of one of them (here $M_B$) is flipped (right), the exchange force results in an increase in the distance between the surface trilayer and the bulk (increase in the distance is exaggerated). **b**, Tunnelling spectra measured for parallel (black; $M_S\uparrow\uparrow M_B$) and antiparallel (purple; $M_S\uparrow\downarrow M_B$) alignment of the magnetizations of the bulk and surface layer. The spectra are normalized (norm.) by the value of $g(V)$ at $V = 8$ mV. Two peaks are observed just below the Fermi level: $P_I$ and $P_{II}$. The energy of peak $P_I(E_{PI})$ depends sensitively on the relative magnetization of the bulk and surface layer. It is 750 μV higher in energy when $M_B$ and $M_S$ are parallel compared with when they are parallel. The energy of peak $P_{II}(E_{PII})$ remains the same. **c**, Change in $M_S$ and $M_B$ observed in the tunnelling spectra taken between 0.41 T and −1.06 T, after having polarized the overall magnetization with $B_z = 1.3$ T. Triangles indicate

$P_I$, as in **b**. The energy of peak $P_I$ jumps abruptly to higher energies when $M_B$ switches between −0.13 T and −0.17 T, resulting in $M_S$ and $M_B$ being antiparallel. The peak jumps back when the surface magnetization switches between −0.96 T and −1.06 T. The spectra are offset vertically for clarity. **d**, Energy $E_{PI}$ of peak $P_I$ as a function of field $B_z$, taken by cycling the field from 1.3 T to −1.06 T and then back to 1.05 T. The direction of the triangles indicates the direction of the ramp of the field. $P_I$ changes abruptly at −0.16 T and −0.98 T when ramping to negative fields, and at 0.20 T and 0.96 T when ramping to a positive field. **e**, Energy $E_{PI}$ as a function of field $B_z$, taken by cycling $B_z$ between −0.25 T and +0.25 T after aligning $M_B$ and $M_S$ with a field of $B_z = -1$ T. All data are taken at $T = 80$ mK (set-point conditions: $V_{set} = 10$ mV, $I_{set} = 450$ pA; lock-in modulation amplitude: $V_L = 125$ μV). The points in **d** and **e** were extracted from a fit of two Lorentzian peaks to the spectra and the error bars are the 95% confidence intervals.The red and blue arrows illustrate the orientation of local moments at the surface (top line) and bulk (bottom line) along the cycle.

increasing field, consistent with its spin-majority character (Supplementary Section 5 and ref. [30]); when the magnetization of the surface layer is antiparallel to the field ($B_B < B_z < B_S$), $P_I$ shifts to higher energies, indicative of a spin-minority character. Once the direction of the surface magnetization has reversed to align with the field, the peak $P_I$ shifts again to lower energies with increasing field ($B_z > 1$ T; Fig. 2c, grey lines), returning to behave as having a spin-majority character. The field-induced shift in the VHS $P_I$ directly confirms the interpretation of the data in Fig. 2a—that of an independent switching of the surface and bulk magnetizations.

Not only can the surface layer be switched independently, but we can also measure the magnetic field loops in which only the bulk magnetization is reversed, by reverting the direction of the field ramp before the surface has switched. Figure 2e shows $P_I$ as a function of magnetic field for a magnetic field loop recorded up to $B_L = 0.25$ T, where $B_L$ is the magnitude of the magnetic field before switching its direction.

The energy difference in the peak position before and after the jump provides an estimate for the magnitude of the exchange coupling experienced by the electronic states associated with the VHS; here this is −750 μeV. From the DFT calculations, we can deduce that energetically as well as from the subtle changes in the tunnelling spectra, it is the full surface triple layer flipping its magnetization, rather than only one RuO₂ layer within the triple layer (Supplementary Section 6 and Supplementary Fig. 8).

## Pinning of surface magnetism

To establish the origin of the different magnetic behaviours of the surface compared with the bulk, we have studied the switching fields at which the surface and bulk magnetizations change as a function of temperature $T$ and of the maximum of the field loop $B_L$. Even though the two peaks in the tunnelling spectrum are very close to each other, necessitating very low temperatures ($T < 500$ mK) to be able to clearly distinguish them, we can trace the changes in the spectra when the relative magnetizations change up to $T \approx 14.3$ K (Fig. 3a–c and Supplementary Section 7). In Fig. 3d–f, we show the representative field loops recorded up to $B_L = +1.3$ T at $T = 2$ K, 6 K and 14.3 K, respectively. The temperature dependence of the fields $B_S$ and $B_B$ at which the surface and bulk switch are summarized in Fig. 3g. We find that (1) the switching field $B_B$ of the bulk has only a weak temperature dependence—consistent with the bulk coercive field $B_B$ obtained from bulk magnetization measurements (Fig. 3g (circles) and Supplementary Fig. 6); and (2) the switching field $B_S$ of the surface has a strong temperature dependence, changing from −1 T at $T = 80$ mK to about 0.2 T at $T = 14.3$ K.

The switching field of the surface also exhibits a notable dependence on the field history, which we discuss here in terms of the loop

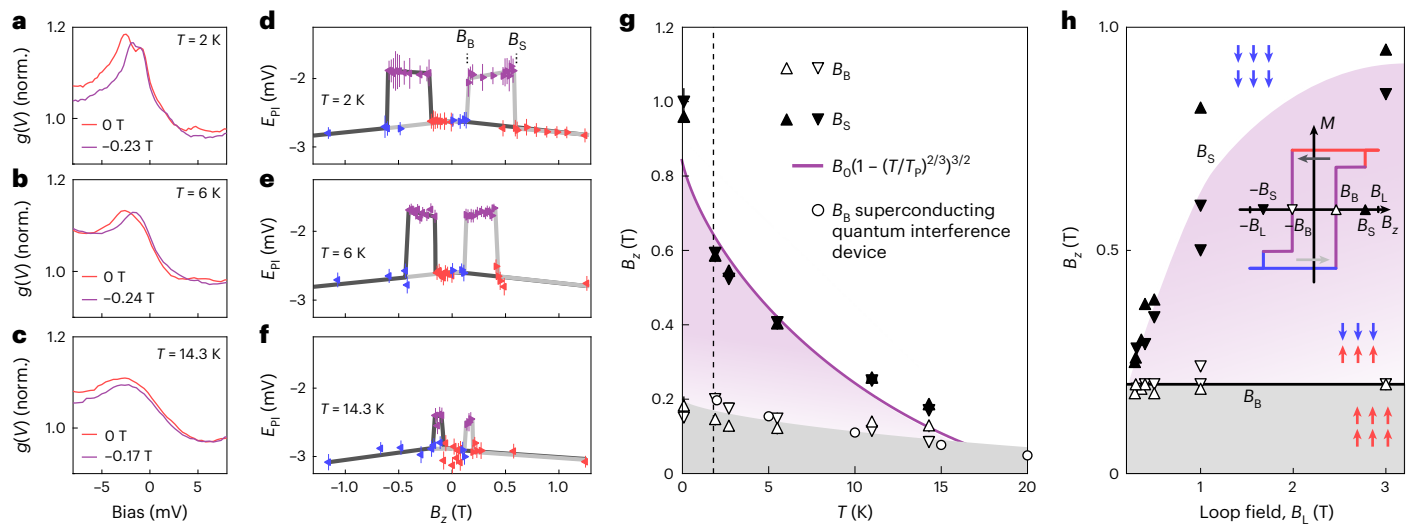

**Fig. 3 | Evidence for domain-wall pinning. a–c,** Tunnelling spectra $g(V)$ acquired at $T$ = 2 K (**a**), 6 K (**b**) and 14.3 K (**c**). The spectra were taken after aligning the sample magnetization at $B_z$ = 1.3 T before recording the spectra at 0 T and a field at which the bulk magnetization has switched. The spectra are normalized by the value of $g(V)$ at 9.8 mV. **d–f,** Energy $E_{Pt}$ of peak $P_1$ as a function of field at $T$ = 2 K (**d**), 6 K (**e**) and 14.3 K (**f**) recorded when cycling the field from $B_z$ = 1.3 T to −1.2 T and back to 1.3 T at $T$ = 2 K (**d**), 6 K (**e**) and 14.3 K (**f**) (points obtained by fitting two Lorentzian functions (**d**) or one Lorentzian function (**e** and **f**); the error bars are 95% confidence intervals). The dashed lines in **d** indicate the switching fields $B_B$ and $B_S$. **g,** Temperature dependence of the fields $B_B$ and $B_S$ at which the bulk (open triangles) and surface (filled triangles) switch magnetization, respectively. The bulk magnetization switches at fields $B_B$ (open triangles); the direction of local moments indicates the points extracted from the upward or downward field ramps (close to the bulk coercive field (open circles; Supplementary Fig. 6), whereas the surface switches

at substantially larger fields. The purple line shows a fit of the coercive field $B = B_0(1 − (T/T_P)^{2/3})^{3/2}$ of a 2D ferromagnet in the presence of strong pinning (Supplementary Section 9; $B_0$ = 0.91 T, $T_P$ = 23 K). The graph shows the field at which $B_S$ and $B_B$ switch following a field ramp looping up to the field ±$B_L$ (indicated on the horizontal axis). **h,** Hysteresis of $B_S$ at $T$ = 1.8 K. The inset shows an illustration of the $M$–$H$ loops performed. $B_S$ varies by more than a factor of two depending on the field history, demonstrating that it exhibits hysteresis, whereas $B_B$ remains constant ($V_{set}$ = 10 mV, $I_{set}$ = 50 pA, except for $B_L$ = 1 T with $I_{set}$ = 100 pA). In **g** and **h**, the white- and grey-shaded areas indicate when the surface and bulk magnetizations are parallel, whereas purple shading indicates when they are antiparallel. The red and blue arrows in **h** show the direction of local moments; the specific direction is indicative only as it depends on whether the ramp started at positive or negative fields.

field $B_L$ to which the field is ramped during each full measurement loop. When changing the field direction immediately after the surface magnetization has switched, the switching field $B_S$ of the surface layer can be trained to lower magnetic fields. Figure 3h shows the fields at which $B_B$ and $B_S$ reverse their magnetizations for different $B_L$ values. This behaviour is fully consistent with what is expected for a ferromagnet due to the randomization of domains and suggests that the magnetic domain structure plays an important role in determining $B_S$. Together with the dramatic temperature dependence of $B_S$ (Fig. 3g) and the large terrace sizes that are frequently larger than 100 nm and sometimes micrometres in lateral size (Supplementary Section 8), these findings suggest that the different magnetic behaviours between the surface layer and the bulk are governed by domain-wall pinning. In the surface layer, the domain wall that triggers the flipping of magnetization is pinned by step edges, a type of defect that does not exist in the bulk and that provides an extended and strong pinning centre. To verify this hypothesis, we have developed a model for the coercive field that accounts for strong pinning in a 2D magnet (Supplementary Section 9). The behaviour expected for the switching field matches the experimentally observed behaviour very well (Fig. 3g, solid purple line), confirming the importance of pinning for the distinct magnetic behaviour of the surface layer.

### Magnetostriction of the surface layer

To establish the structural changes due to the exchange force associated with the flip of the magnetization direction, we record the change in the vertical tip position $\Delta z$ as the magnetic field is ramped. Such traces have been used previously to study magnetostriction[29,32]. Here we search for sudden changes when the relative magnetization of the surface layer and the bulk changes. Figure 4a,b shows two traces of the vertical tip position $z(H)$ as the magnetic field $B_z$ is ramped across the field $B_S$ at which the surface layer switches. The magnetostriction curves exhibit jumps of less than 500 fm around the field at which the magnetization reversal of the surface layer occurs from antiparallel to parallel to the bulk magnetization. To verify this assignment, we have recorded the tip height $z(t)$ as a function of time $t$ as the field is ramped in steps of 10 mT and checking the orientation of magnetization of the surface layer before and after the ramp from the tunnelling spectra (Supplementary Section 10). Figure 4c,d shows two $z(t)$ traces recorded when the surface layer had switched. Both traces reveal a jump of the same order. Since the only change between the tunnelling spectra is the change in the field and the change in the peak in the spectra (as shown in the corresponding insets), which is consistent with the shift in $P_1$ seen at lower temperatures, we can assign the jump to the surface layer switching its magnetization direction. We note that the switching does not always occur during ramping the field, but sometimes before (Fig. 4c) or after (Fig. 4d). This can be ascribed to the activation of the depinning of the domain wall, which is a statistical process. Figure 4e shows a histogram of such jumps in the magnetostriction curves, which show a relatively sharp peak close to 500 fm and which we attribute to the magnetostriction of the surface layer.

## Discussion

The ability to control the magnetization of the surface layer independently of that of the bulk opens up an opportunity to study the interplay between crystal structure, low-energy electronic states and magnetism in a strongly correlated electron material in unprecedented detail, providing a reference point for understanding and modelling the coupling between electronic and lattice degrees of freedom in these materials. The subtle changes in the tunnelling spectra for parallel and antiparallel magnetization of the surface and the bulk provide clues for the nature of the coupling between the surface and subsurface layer. The spectrum shows a peak shifting by about 750 μeV. In a minimal model that describes the $Sr_4Ru_3O_{10}$ layer by just one of the three $RuO_2$ layers, this change can be rationalized by the hybridization between the spin–orbit-coupling-induced VHSs between adjacent layers: in the antiferromagnetic case, this hybridization is zero (or close to zero if the spins are not perfectly collinear), whereas in the ferromagnetic case, it is non-zero, resulting in a splitting (Supplementary Fig. 5). Using previously determined parameters of the band structure[30], this coupling energy is close to the value of 750 μeV found experimentally. Although spin–orbit coupling plays an important role in the electronic structure, the agreement found from our minimal model suggests that it (and hence the magnetic anisotropy) affects the magnetism of the surface and bulk in the same way, and, consequently, plays only a minor role in the difference in their magnetic properties—as is expected if the spins remain mostly aligned along the crystallographic $c$ direction. This is confirmed by the field dependence of the energy of the VHS having a slope of a constant magnitude independent of the field strength and direction, and the absence of Rashba spin splitting in the electronic structure of the surfaces of ruthenates[30,33]. For example, a non-negligible in-plane component of the magnetization would be expected to result in anisotropic QPI patterns as observed in $Sr_3Ru_2O_7$ (ref. 13) as well as in characteristic changes in the tunnelling spectra when the magnetization aligns with the field, which are both not observed in $Sr_4Ru_3O_{10}$ (ref. 30). Given the success of our model for the coercive field of a 2D magnet, differences in the pinning between surface and bulk are a much more likely explanation.

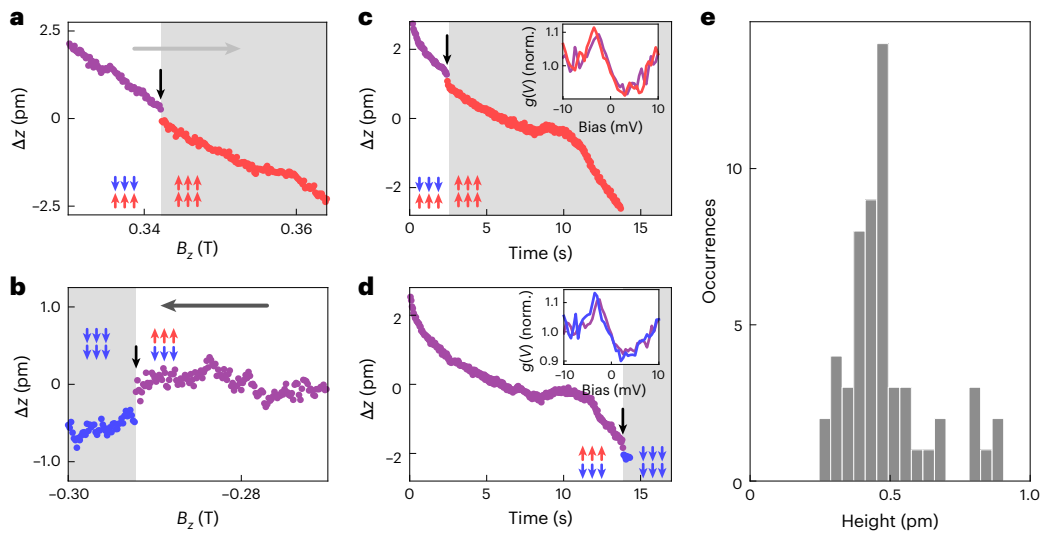

**Fig. 4 | Magnetostriction of the surface layer. a,b,** Change in $\Delta z(H)$ recorded as the field $B_z$ is ramped from −0.4 T to 0.4 T (**a**) and back (**b**) ($T = 1.8$ K, $V_{set} = 10$ mV, $I_{set} = 50$ pA). Only a small section of the ramp is shown; the full ramps are shown in Supplementary Fig. 14b. The traces show a small jump when the surface layer switches its magnetization, indicated by the black arrow. **c,d,** Time traces of the $\Delta z(t)$ over –15 s during which the field $B_z$ is ramped by 10 mT from 0.34 T to 0.35 T (**c**) and from −0.27 T to −0.28 T (**d**). A jump of −300 fm can be seen in both traces, marked by a black arrow. The grey-shaded area indicates the time after the switch has happened. The switching of the surface layer is verified from the tunnelling spectra taken before (purple line) and after (red/blue line) the time trace is recorded (insets show the tunnelling parameters for $z(H)$ and $z(t)$ traces, $V_{set} = 10$ mV, $I_{set} = 50$ pA; for the spectra, $I_{set} = 450$ pA, $V_L = 0.5$ mV; Supplementary Fig. 14j,k). **e,** Histogram of jumps in $\Delta z(t)$ when the surface layer switches, extracted from 56 magnetostriction curves ($I_{set} = 50$ pA, $V_{set} = 10$–50 mV).

The most striking finding is that we can detect a clear change in the height of the surface, by about 500 fm, when the magnetization of the surface layer is switching. It suggests that the surface layer contracts slightly when its magnetization switches from being aligned antiparallel with the bulk to parallel, consistent with DFT calculations (Supplementary Fig. 2 and Supplementary Section 2). Although the change recorded in STM could, in principle, be a consequence of the apparent height and changes in the electronic structure, the jump we observe is largely independent of the bias voltage (Supplementary Fig. 18), and the changes in the electronic structure are extremely subtle (Supplementary Fig. 8a,b). We can also exclude that this change in apparent height is a consequence of a change in the work function of the sample (Supplementary Section 10). The height change of $\Delta z = 500$ fm corresponds to a relative height change of $\Delta l_r = \frac{\Delta z}{c/2} = 3.5 \times 10^{-4} = 350$ ppm

This value of magnetostriction is giant for an exchange-driven mechanism in a transition metal compound. Similarly large magnetostrictions are only observed in rare earth compounds (for example, ref. 34) due to metamagnetic transitions[35] or with the reorientation of magnetic moments and when spin–orbit coupling is important[23,36]. Our spectroscopic data show that spin–orbit coupling plays only a minor role here, enabling the identification of the exchange-driven mechanism.

The relaxation of the surface layer can be rationalized from the Bethe–Slater curve shown in Fig. 1c: switching between parallel and antiparallel alignment of the surface and the bulk, the distance dependence of the exchange interaction results in an exchange force that acts on the surface layer, and pushes it away from the bulk for antiparallel alignment. Although the force and this effect are expected to be tiny, they provide a natural explanation for the relaxation we observe here. A strong magnetoelastic coupling in $Sr_4Ru_3O_{10}$, as suggested by these measurements, is also found in experiments under hydrostatic pressure[27,37] and in dilatometric measurements[28]. Our experiments provide microscopic insights into the mechanism and demonstrate the important and possibly dominant role of the exchange force for this giant magnetoelastic coupling. The spin–orbit-coupling-mediated VHSs in the vicinity of the Fermi energy[30] provide a route to detect the relative magnetizations, and may contribute in driving the dramatic response of the lattice to the change in magnetization. The magnitude of the structural change that we detect is much larger than what is obtained from DFT (Supplementary Section 2). DFT calculations for this system confirm two key findings: (1) for a ferromagnetic configuration, the interlayer distance is slightly smaller than for an antiferromagnetic configuration; (2) a calculation of the exchange energy as a function of layer separation for two magnetic layers of $Sr_2RuO_4$ (with octahedral rotations) indeed confirms a Bethe–Slater-like behaviour, as our magnetostriction measurements suggest (Supplementary Fig. 2).

Although DFT calculations generally appear to capture the qualitative features of the electronic structure of ruthenates fairly well (for example, refs. 30,33,38–42), the height change due to exchange magneto-striction found in DFT is almost an order of magnitude smaller than the one found experimentally. This is probably due to the shortcomings of DFT to capture the electronic correlation effects in this material, which is also reflected in the band renormalization between the band structure obtained from DFT and the one determined by angle-resolved photoemission spectroscopy and quasi-particle interference[30,41].

## Conclusion

Our results demonstrate a direct link between exchange interaction, electronic structure and crystal structure, providing a microscopic insight into the structure–property relationships in a metamagnetic quantum material. The understanding that can be derived from this model system provides experimental results on which to test the theoretical descriptions of strongly correlated electron materials, particularly in modelling of the coupling between electronic and lattice degrees of freedom. We find substantially larger magneto-structural coupling than expected from DFT calculations, suggesting that electronic correlation effects play an important role. Materials modelling that can properly capture the correlation-enhanced coupling between the electronic and structural degrees of freedom promises routes to tailor this coupling and exploit it, for example, in the stabilization of new correlated ground states and for unconventional superconductivity. The 2D nature of magnetism in $Sr_4Ru_3O_{10}$ makes this material an ideal system to study the influence of dimensionality on the magnetic properties and establish the implications of strong magnetostructural coupling. A detailed understanding of the interplay between magnetism, spin–orbit coupling and crystal structure paves the way to fine-tune the properties of these materials, enabling field control of the electronic structure.[12–14] Our experiments demonstrate novel routes to the purely electronic or structural read out of magnetic information, using VHSs as probes of the exchange interaction or by exploiting the giant magnetostructural coupling—demonstrated here at the level of a single domain wall between the surface and subsurface layers.

## Online content

Any methods, additional references, Nature Portfolio reporting summaries, source data, extended data, supplementary information, acknowledgements, peer review information; details of author contributions and competing interests; and statements of data and code availability are available at https://doi.org/10.1038/s41567-025-02893-x.

## References

1.   Tranquada, J. M., Sternlieb, B. J., Axe, J. D., Nakamura, Y. & Uchida, S. Evidence for stripe correlations of spins and holes in copper oxide superconductors. *Nature* **375**, 561–563 (1995).

2.   Howald, C., Eisaki, H., Kaneko, N. & Kapitulnik, A. Coexistence of periodic modulation of quasiparticle states and superconductivity in $Bi_2Sr_2CaCu_2O_{8+\delta}$. *Proc. Natl Acad. Sci. USA* **100**, 9705–9709 (2003).

3.   Hanaguri, T. et al. A 'checkerboard' electronic crystal state in lightly hole-doped $Ca_{2-x}Na_xCuO_2Cl_2$. *Nature* **430**, 1001–1005 (2004).

4.   Kohsaka, Y. et al. How Cooper pairs vanish approaching the Mott insulator in $Bi_2Sr_2CaCu_2O_{8+\delta}$. *Nature* **454**, 1072–1078 (2008).

5.   Ghiringhelli, G. et al. Long-range incommensurate charge fluctuations in $(Y,Nd)Ba_2Cu_3O_{6+x}$. *Science* **337**, 821–825 (2012).

6.   Du, Z. et al. Periodic atomic displacements and visualization of the electron-lattice interaction in the cuprate. *Phys. Rev. X* **13**, 021025 (2023).

7.   Ramirez, A. P. Colossal magnetoresistance. *J. Phys.: Condens. Matter* **9**, 8171–8199 (1997).

8.   Ohmichi, E., Yoshida, Y., Ikeda, S. I., Shirakawa, N. & Osada, T. Colossal magnetoresistance accompanying a structural transition in a highly two-dimensional metallic state of $Ca_3Ru_2O_7$. *Phys. Rev. B* **70**, 104414 (2004).

9.   Lin, X. N., Zhou, Z. X., Durairaj, V., Schlottmann, P. & Cao, G. Colossal magnetoresistance by avoiding a ferromagnetic state in the Mott system $Ca_3Ru_2O_7$. *Phys. Rev. Lett.* **95**, 017203 (2005).

10.  Satoh, H. & Ohkawa, F. J. Theory of the metamagnetic crossover in $CeRu_2Si_2$. *Phys. Rev. B* **63**, 184401 (2001).

11.  Gegenwart, P., Weickert, F., Perry, R. S. & Maeno, Y. Low-temperature magnetostriction of $Sr_3Ru_2O_7$. *Phys. B: Condens. Matter* **378**, 117–118 (2006).

12.  Yin, J.-X. et al. Giant and anisotropic many-body spin-orbit tunability in a strongly correlated kagome magnet. *Nature* **562**, 91–95 (2018).

13.  Naritsuka, M. et al. Compass-like manipulation of electronic nematicity in $Sr_3Ru_2O_7$. *Proc. Natl Acad. Sci. USA* **120**, e2308972120 (2023).

14.  Cheng, E. et al. Tunable positions of Weyl nodes via magnetism and pressure in the ferromagnetic Weyl semimetal CeAlSi. *Nat. Commun.* **15**, 1467 (2024).

15.  Slater, J. C. Atomic shielding constants. *Phys. Rev.* **36**, 57–64 (1930).

16.  Sommerfeld, A. & Bethe, H. in *Elektronentheorie der Metalle* 333–622 (Springer, 1933).

17.  Azumi, K. & Goldman, J. E. Volume magnetostriction in nickel and the Bethe-Slater interaction curve. *Phys. Rev.* **93**, 630 (1954).

18.  Mokrousov, Y., Bihlmayer, G., Blügel, S. & Heinze, S. Magnetic order and exchange interactions in monoatomic 3*d* transition-metal chains. *Phys. Rev. B* **75**, 104413 (2007).

19.  Gallagher, K. A., Willard, M. A., Zabenkin, V. N., Laughlin, D. E. & McHenry, M. E. Distributed exchange interactions and temperature dependent magnetization in amorphous $Fe_{88-x}Co_xZr_7B_4Cu_1$ alloys. *J. Appl. Phys.* **85**, 5130–5132 (1999).

20.  Wang, C. et al. Bethe-Slater-curve-like behavior and interlayer spin-exchange coupling mechanisms in two-dimensional magnetic bilayers. *Phys. Rev. B* **102**, 020402 (2020).

21.  Jiang, S., Xie, H., Shan, J. & Mak, K. F. Exchange magnetostriction in two-dimensional antiferromagnets. *Nat. Mater.* **19**, 1295–1299 (2020).

22.  Chen, W. et al. Direct observation of van der Waals stacking-dependent interlayer magnetism. *Science* **366**, 983–987 (2019).

23.  Sander, D. *Handbook of Magnetism and Magnetic Materials* (eds Coey, J. M. D. & Parkin, S. S. P.) 549–593 (Springer, 2021).

24.  Cao, G. et al. Competing ground states in triple-layered $Sr_4Ru_3O_{10}$: verging on itinerant ferromagnetism with critical fluctuations. *Phys. Rev. B* **68**, 174409 (2003).

25.  Lin, X. N., Bondarenko, V. A., Cao, G. & Brill, J. W. Specific heat of $Sr_4Ru_3O_{10}$. *Solid State Commun.* **130**, 151–154 (2004).

26.  Crawford, M. K. et al. Structure and magnetism of single crystal $Sr_4Ru_3O_{10}$: a ferromagnetic triple-layer ruthenate. *Phys. Rev. B* **65**, 214412 (2002).

27.  Zheng, H. et al. Observation of a pressure-induced transition from interlayer ferromagnetism to intralayer antiferromagnetism in $Sr_4Ru_3O_{10}$. *Phys. Rev. B* **98**, 064418 (2018).

28.  Schottenhamel, W. et al. Dilatometric study of the metamagnetic and ferromagnetic phases in the triple-layered $Sr_4Ru_3O_{10}$ system. *Phys. Rev. B* **94**, 155154 (2016).

29.  Benedičič, I. et al. Interplay of ferromagnetism and spin-orbit coupling in $Sr_4Ru_3O_{10}$. *Phys. Rev. B* **106**, L241107 (2022).

30.  Marques, C. A. et al. Spin-orbit coupling induced Van Hove singularity in proximity to a Lifshitz transition in $Sr_4Ru_3O_{10}$. *npj Quantum Mater.* **9**, 35 (2024).

31.  Meier, F., Zhou, L., Wiebe, J. & Wiesendanger, R. Revealing magnetic interactions from single-atom magnetization curves. *Science* **320**, 82–86 (2008).

32. Trainer, C., Abel, C., Bud'ko, S. L., Canfield, P. C. & Wahl, P. Phase diagram of $CeSb_2$ from magnetostriction and magnetization measurements: evidence for ferrimagnetic and antiferromagnetic states. *Phys. Rev. B* **104**, 205134 (2021).

33. Tamai, A. et al. High-resolution photoemission on $Sr_2RuO_4$ reveals correlation-enhanced effective spin-orbit coupling and dominantly local self-energies. *Phys. Rev. X* **9**, 021048 (2019).

34. Gratz, E., Lindbaum, A., Markosyan, A. S., Mueller, H. & Yu Sokolov, A. Isotropic and anisotropic magnetoelastic interactions in heavy and light $RCo_2$ laves phase compounds. *J. Phys.: Condens. Matter* **6**, 6699 (1994).

35. Szymczak, H. Giant magnetostrictive effects in magnetic oxides. *J. Magn. Magn. Mater.* **40**, 066501 (2000).

36. He, J.-C. et al. Magnetic-field-induced sign changes of thermal expansion in $DyCrO_4$. *Chin. Phys. Lett.* **40**, 066501 (2023).

37. Gupta, R., Kim, M., Barath, H., Cooper, S. L. & Cao, G. Field- and pressure-induced phases in $Sr_4Ru_3O_{10}$: a spectroscopic investigation. *Phys. Rev. Lett.* **96**, 067004 (2006).

38. Veenstra, C. N. et al. Determining the surface-to-bulk progression in the normal-state electronic structure of $Sr_2RuO_4$ by angle-resolved photoemission and density functional theory. *Phys. Rev. Lett.* **110**, 097004 (2013).

39. Kreisel, A. et al. Quasi-particle interference of the Van Hove singularity in $Sr_2RuO_4$. *npj Quantum Mater.* **6**, 100 (2021).

40. Gebreyesus, G. et al. Electronic structure and magnetism of the triple-layered ruthenate $Sr_4Ru_3O_{10}$. *Phys. Rev. B* **105**, 165119 (2022).

41. Ngabonziza, P. et al. Layer-dependent spin-resolved electronic structure of ferromagnetic triple-layered ruthenate $Sr_4Ru_3O_{10}$. *Phys. Rev. B* **111**, 115146 (2025).

42. Chandrasekaran, A. et al. On the engineering of higher-order Van Hove singularities in two dimensions. *Nat. Commun.* **15**, 9521 (2024).

## Methods

### Sample growth

The single-crystal samples of $Sr_4Ru_3O_{10}$ used in this study are from the same batch of samples as those measured in refs. 29,30. The growth procedure is described in detail in ref. 43. The single crystals were grown using Ru self-flux by the floating-zone method. Feed rods were prepared by a standard solid-state reaction from repeated thermal cycles of a mixture of $SrCO_3$ (99.99%) and $RuO_2$ (99.9% purity). An excess of $RuO_2$ was added to account for the evaporation of Ru during growth. For the STM measurements, the samples were shaped in small rectangular pieces with an average size of $1 \times 1 \times 0.3$ mm³, with the crystallographic $c$ axis normal to the plate-like shape.

### Sample characterization

The quality of the grown single crystals was determined by X-ray diffraction, energy- and wavelength-dispersive X-ray spectroscopy, and electron backscattered diffraction. We have measured the bulk magnetization of single crystals of $Sr_4Ru_3O_{10}$ from the same batch as the crystals used for the STM measurements using the vibrating-sample magnetometry option of a physical property measurement system by Quantum Design. This was used to confirm the magnetic properties of the samples, as well as to determine the coercive field as a function of temperature (Supplementary Section 4).

### STM

STM and scanning tunnelling spectroscopy measurements have been acquired in two instruments, one mounted in a dilution refrigerator operating at temperatures down to below 100 mK and in magnetic fields up to 14 T (ref. 44), and one mounted in a custom-built insert with a 1-K pot operating at temperatures down to 1.8 K and fields up to 14 T (ref. 45). In both systems, the field $B_z$ is applied normal to the sample surface. The bias voltage $V$ is applied to the sample, with the tip at virtual ground. Tunnelling spectra are acquired in open-feedback loop conditions using a lock-in technique to detect the differential conductance, with the lock-in modulation $V_L$ added to the bias voltage $V$. All measurements were done with a non-magnetic tip. The tip was cut from a PtIr wire and prepared in situ by field emission on a gold target.

The spectra shown in Figs. 2 and 3 were obtained by averaging an $8 \times 8$ grid of spectra acquired over the same $(1 \times 1)$ nm² area, using $V_{set} = 10$ mV and $I_{set} = 450$ pA as the set-point parameters. Lock-in modulations were $V_L = 400$ μV at 14.3 K, 10 K and 6 K, $V_L = 250$ μV at 2 K and $V_L = 125$ μV at 80 mK. Data in Fig. 4 were acquired at $T = 1.8$ K and using $V_L = 500$ μV.

The error bars were obtained from fits and represent 95% confidence intervals (Figs. 2d,e and 3d–g and Supplementary Figs. 7b and 9g–o). In Fig. 3g and Supplementary Fig. 7b, they are smaller than the symbols.

Magnetostriction curves shown in Fig. 4 were acquired in the closed-feedback-loop condition, recording $z(H)$ (or $z(t)$) as the magnetic field is ramped. To minimize vibrational noise, for the curves shown in Fig. 4, we have trained the field at which the surface layer changes magnetization to a field between 0.3 T and 0.4 T. Supplementary Section 10 and Supplementary Fig. 14 provide additional information about the magnetostriction measurements.

Magnetic fields $B_z$ quoted throughout the text refer to the field in vacuum, $B_z = \mu_0 H_z$, in the centre of the superconducting magnet.

### DFT calculations

To confirm the structural changes for different magnetic configurations, we have carried out extensive DFT calculations. The calculations have been performed using QUANTUM ESPRESSO[46] and Vienna ab initio simulation package[47–52]. We have performed structural relaxations of surface slabs as well as bilayers of $Sr_4Ru_3O_{10}$ using Perdew–Burke–Ernzerhof as the exchange–correlation functional. Calculations in the Vienna ab initio simulation package have been done using $8 \times 8 \times 1$ k points and with a plane wave cut-off of 1,700 eV. The results have been cross-checked using PBESol and using LDA + $U$. We consistently find that the ferromagnetic configuration has a 50–100-fm-smaller inter-layer separation compared with the antiferromagnetic configuration. We have also performed calculations for two layers of $Sr_2RuO_4$ including octahedral rotations to study the exchange interaction between the layers as a function of layer separation and for a minimal model to understand the changes in the electronic structure (Supplementary Section 2). We find a Bethe–Slater-type distance dependence of the exchange interaction (Supplementary Fig. 2) from these calculations. In previous work, we have demonstrated that the electronic structure of a single layer of $Sr_2RuO_4$ with octahedral rotations provides a good approximation for the low-energy electronic structure of $Sr_4Ru_3O_{10}$ (ref. 30). The results for the magnetostriction obtained for calculations for two layers of $Sr_2RuO_4$ are also consistent with those for $Sr_4Ru_3O_{10}$.

## Data availability

## References

43. Fittipaldi, R., Sisti, D., Vecchione, A. & Pace, S. Crystal growth of a lamellar $Sr_3Ru_2O_7$–$Sr_4Ru_3O_{10}$ eutectic system. *Cryst. Growth Des.* **7**, 2495–2499 (2007).

44. Singh, U. R., Enayat, M., White, S. C. & Wahl, P. Construction and performance of a dilution-refrigerator based spectroscopic-imaging scanning tunneling microscope. *Rev. Sci. Instrum.* **84**, 013708 (2013).

45. White, S. C., Singh, U. R. & Wahl, P. A stiff scanning tunneling microscopy head for measurement at low temperatures and in high magnetic fields. *Rev. Sci. Instrum.* **82**, 113708 (2011).

46. Giannozzi, P. et al. Advanced capabilities for materials modelling with Quantum ESPRESSO. *J. Phys.: Condens. Matter* **29**, 465901 (2017).

47. Kresse, G. & Hafner, J. Ab initio molecular dynamics for liquid metals. *Phys. Rev. B* **47**, 558–561 (1993).

48. Kresse, G. & Hafner, J. Ab initio molecular-dynamics simulation of the liquid-metal–amorphous-semiconductor transition in germanium. *Phys. Rev. B* **49**, 14251–14269 (1994).

49. Kresse, G. & Hafner, J. Norm-conserving and ultrasoft pseudopotentials for first-row and transition elements. *J. Phys.: Condens. Matter* **6**, 8245–8257 (1994).

50. Kresse, G. & Furthmüller, J. Efficiency of ab-initio total energy calculations for metals and semiconductors using a plane-wave basis set. *Comput. Mater. Sci.* **6**, 15–50 (1996).

51. Kresse, G. & Furthmüller, J. Efficient iterative schemes for ab initio total-energy calculations using a plane-wave basis set. *Phys. Rev. B* **54**, 11169–11186 (1996).

52. Kresse, G. & Joubert, D. From ultrasoft pseudopotentials to the projector augmented-wave method. *Phys. Rev. B* **59**, 1758–1775 (1999).

53. Marques, C. A. et al. Exchange-driven giant magnetoelastic coupling in a correlated itinerant ferromagnet (dataset). *University of St Andrews Research Portal* https://doi.org/10.17630/be94879c-bfcc-460f-9061-764b1212d943 (2025).

## Acknowledgements

We gratefully acknowledge discussions with S.-W. Cheong, P. Hirschfeld, P. Littlewood, C. Stock and S. White. C.A.M., L.C.R., M.N. and P.W. gratefully acknowledge funding from the Engineering and Physical Sciences Research Council through EP/R031924/1, EP/S005005/1 and EP/X015556/1 and from the Leverhulme Trust through Research Project grant RPG-2022-315, I.B. through the International Max Planck Research School for Chemistry and Physics of Quantum Materials, and H.L. from a fellowship from the Royal Commission of the Exhibition of 1851. C.A.M. was supported by the Federal Commission for Scholarships for Foreign Students for the Swiss Government Excellence Scholarship (ESKAS no. 2023.0017) for the academic year 2023–24. R.F., M.L. and A.V. thank the EU's Horizon 2020 research and innovation programme under grant agreement no. 964398 (SUPERGATE). This work used computational resources of the Cirrus UK National Tier-2 HPC Service at EPCC funded by the University of Edinburgh and EPSRC (EP/P020267/1) and of the High-Performance Computing cluster Kennedy of the University of St Andrews.

## Author contributions

## Competing interests

The authors declare no competing interests.

## Additional information

**Supplementary information** The online version contains supplementary material available at https://doi.org/10.1038/s41567-025-02893-x.

**Correspondence and requests for materials** should be addressed to Carolina A. Marques or Peter Wahl.

**Peer review information** *Nature Physics* thanks Jia-Xin Yin and the other, anonymous, reviewer(s) for their contribution to the peer review of this work.

**Reprints and permissions information** is available at www.nature.com/reprints.